\documentclass{article}
\usepackage{spconf,amsmath,graphicx}
\usepackage{hyperref}
\usepackage{amsmath}
\usepackage{dsfont}
\usepackage{setspace}

\title{A Two-Stage Approach to Device-Robust Acoustic Scene Classification}

\name{
\begin{tabular}{@{}c@{}}
Hu Hu$^{1}$,
      Chao-Han Huck Yang$^{1}$,
      Xianjun Xia$^{2}$,
      Xue Bai$^{3}$,
      Xin Tang$^{3}$, \\
      Yajian Wang$^{3}$,
      Shutong Niu$^{3}$,
      Li Chai$^{3}$,
      Juanjuan Li$^{2}$,
      Hongning Zhu$^{2}$,
      Feng Bao$^{2}$, \\
      Yuanjun Zhao$^{2}$,
      Sabato Marco Siniscalchi$^{1, 4}$,
      Yannan Wang$^{2}$,
      Jun Du$^{3}$,
      Chin-Hui Lee$^{1}$
      \end{tabular}}

\address{$^1$School of Electrical and Computer Engineering, Georgia Institute of Technology, GA, USA \\
$^2$Tencent Media Lab, Tencent Corporation, China\\
$^3$University of Science and Technology of China, HeFei, China\\
$^4$Computer Engineering School, University of Enna Kore, Italy\\
\{huhu, huckiyang, chinhui.lee\}@gatech.edu, yannanwang@tencent.com, jundu@ustc.edu.cn}

\begin{document}
\ninept
\maketitle
\begin{abstract}
To improve device robustness, a highly desirable key feature of a competitive data-driven acoustic scene classification (ASC) system,
a novel two-stage system based on fully convolutional neural networks (CNNs) is proposed. Our two-stage system leverages on an ad-hoc score combination based on two CNN classifiers: (i) the first CNN classifies acoustic inputs into one of three broad classes, and (ii) the second CNN classifies the same inputs into one of ten finer-grained classes. Three different CNN architectures are explored to implement the two-stage classifiers, and a frequency sub-sampling scheme is investigated. Moreover,  novel data augmentation schemes for ASC are also investigated. Evaluated on DCASE 2020 Task 1a, our results show that the proposed ASC system attains a state-of-the-art accuracy on the development set, where our best system, a two-stage fusion of CNN ensembles, delivers a 81.9\% average accuracy among multi-device test data, and it obtains a significant improvement on unseen devices. Finally, neural saliency analysis with class activation mapping (CAM) gives new insights on the patterns learnt by our models.
\end{abstract}
\section{Introduction}
\label{sec:intro}
Acoustic scene classification (ASC) refers to the task of identifying real-life sounds into environment classes, such as metro station, street traffic, public square, etc. An acoustic scene sound contains much information and rich content, which makes accurate scene prediction difficult and an intriguing research problem at the same time. In recent years, the organizers of the Detection and Classification of Acoustic Scenes and Events (DCASE) challenge \cite{dcase2017, dcase2018, dcase2020} provided both the benchmark data and a competitive platform to promote acoustic scene research and analysis. If we are to analyze top ASC systems reported in the challenge, we will find that most of them are built on the deep neural networks (DNNs) framework, and the key ingredient of their success is the use of convolutional layers \cite{asc-cnn1, asc-cnn2, asc-cnn3, dcase-2020-rank2, asc-robust}. Advanced deep learning techniques, such as attention mechanism \cite{asc-attention1, asc-attention2}, mix-up \cite{asc-mixup2, asc-mixup1}, generative adversarial network (GAN) and variational auto encoder (VAE) based data augmentation  \cite{asc-gan1, asc-gan2}, and deep feature learning \cite{openl3, asc-feature1, asc-feature2} can further enhance ASC results. Although those ASC systems work well on in-domain data, a generalization issue is observed in the presence of audio recording collected in mismatched testing conditions, e.g., audio segments recorded with devices different from those used in the training phase \cite{dcase2018, dcase2020}. Device robustness is an inevitable issue in a real production environment, and it is an important aspect to handle when designing an ASC system.

In this work, we have three main contributions: (i) we propose a novel two-stage ASC system;
(ii) we investigate three different fully CNN models in the proposed two-stage system;
(iii) we explore novel data augmentation strategies to reduce the device dependency of our models. Our experimental results are gathered on the DCASE 2020 task1a development data set, and the proposed CNN models achieve competitive results. Specifically, we attained a 79.4\% overall ASC accuracy by ensemble of  our three CNN models, and it obtains a significant improvement on unseen devices when compared with the baseline. The two-stage approach further boosts the accuracy of the ensemble model, and an ASC classification accuracy of 81.9\% is attained. To better understand the key characteristics of our CNN based systems, we carry out a neural saliency analysis based on the class activation mapping (CAM) \cite{cam} on input data and the final output. Such analysis shows that our CNN models actually focus more on the meaningful audio segments containing acoustic events for the final classification task, such as bird sounds in a park or car horn sounds from street traffic, rather than leveraging on the background sound, as instead reported in \cite{asc-cam}.

\begin{figure}[t]
  \centering
  \includegraphics[width=0.75\linewidth]{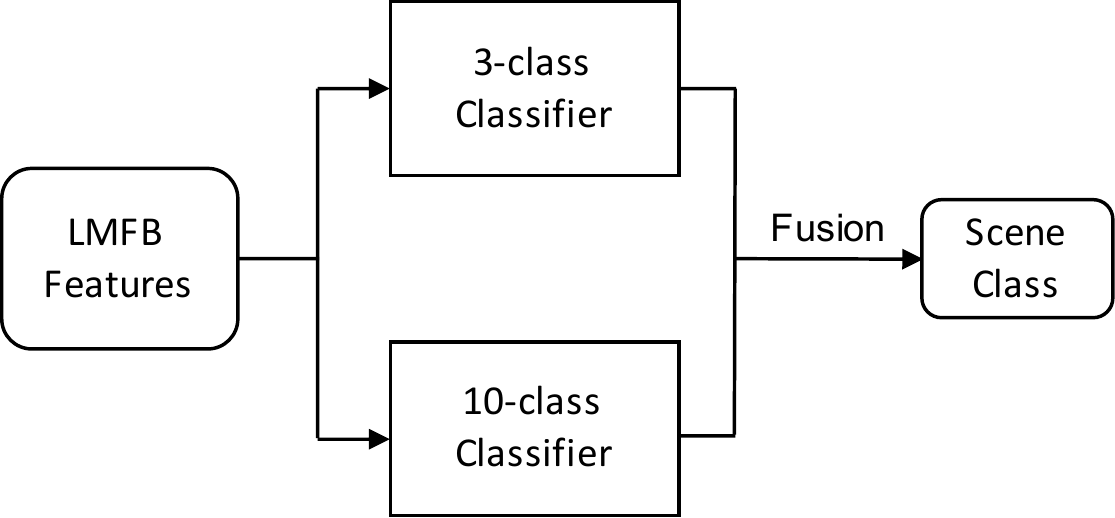}
  \caption{The proposed two-stage ASC system.}
  \label{fig:system}
  \vspace{-0.5cm}
\end{figure}

\vspace{-0.0cm}
\section{Two-Stage ASC System Design}
\label{sec:sys}

\vspace{-0.1cm}
\subsection{Two-Stage Classification Procedure}
\vspace{-0.1cm}

Figure~\ref{fig:system} shows the proposed two-stage ASC system. The overall system consists of two independent classifiers and outputs the class of the input audio scene choosing among ten classes. Different from a more conventional one-stage ASC systems, which directly feed acoustic features into the 10-class classifier and get final scene class, an extra general classifier, namely a 3-class classifier is introduced, which we expect it to enhance the classification process and leverage the over-fitting issue of 10-class classifier.

In our setup, the 3-class classifier classifies an input scene audio into one of three broad classes: in-door, out-door, and transportation. This 3-class classification way is from our prior knowledge that scene audios can be roughly categorized into such three classes. The 10-class classifier is actually the main classifier, which assigns a given input audio clip into one of ten target acoustic scene classes, including airport, shopping mall, metro station, pedestrian street, public square, street traffic, tram, bus, metro, park. Each audio clip should belong to one  of the three / ten classes. The final acoustic scene class is chosen by the score fusion of those two classifiers. If we let $\mathds{C}^1$ and $\mathds{C}^2$  denote the set of three broad classes, and ten classes, respectively, and let $F^1$ and $F^2$ indicate the output of the first and second classifier, respectively. The final predicted class $Class(x)$ for the input $x$ is:
\begin{equation}
    Class(x) = \underset{q, (p\in \mathds{C}^1, q\in \mathds{C}^2, p \supset q)}{\operatorname{argmax}}\ F^1_p(x) * F^2_{q}(x), \nonumber
\end{equation}
\noindent where $p \supset q$ means that $p$ can be thought of a super set of $q$. For example, transportation class is the super set for bus, tram, and metro classes. Therefore, the probability of an input audio clip to be from the public square scene is equal to the product  of the probability of out-door place, $F^1_p(x)$, and that of public square, $F^2_q(x)$.

\vspace{-0.2cm}
\subsection{CNN based Classifiers and Ensemble}
\vspace{-0.1cm}
Three CNN based models are investigated to build our ASC system, namely Resnet, FCNN and fsFCNN. Resnet is a deep residual neural network \cite{resnet} empowered by residual learning. Residual learning is a major efficient method to avoid gradient vanish issues while increasing the depth of networks to allow high-level feature learning. Our Resnet structure is based on the model in \cite{dcase-resnet}, which has no frequency sub-sampling throughout the whole network. Each input feature map is divided into two sub-feature mapping along the frequency dimension. To be specific, if we have $N$ frequency bins, the first $N/2$ and the second half are processed by two parallel stacked residual layers. A global pooling layer and 10-way soft-max are used to get the final utterance level prediction results. 

The FCNN (fully convolutional neural network) model is a VGG \cite{vgg}-like model which is built with 9 stacked convolutional layers with small-size kernel. Each convolutional layer is followed by a batch normalization operation and ReLU activation function. Dropout is also used in order to alleviate over-fitting issues. Before the final global average pooling layer, channel attention \cite{bi2020multiple} is applied to each output channel of the last layer. fsFCNN (frequency sub-sampling FCNN) is an extension of FCNN model which mainly has 2 more convolutional layers and reduces max-pooling size in the frequency axis. In our experiments we notice that it can help to leverage over-fitting issue.

Finally, we build an ensemble using those CNN models and make the final predictions. The output probabilities after softmax are summed by equal weights, and then the scene class with the maximum posterior represents the final classification decision. For the 3-class classifier, only Resnet and FCNN are used; whereas, for the 10-class classifier, all three CNN models are used. The two-stage score fusion is performed after model ensemble of classifiers in each stage.

\vspace{-0.1cm}
\subsection{Data Augmentation Strategies for Training}
\vspace{-0.1cm}
When training the CNN models, 9 different data augmentation methods are investigated. The training for 3-class models and 10-class models adopt the same data augmentation strategy. They can be categorized into two main classes: generating extra data or not.

Data augmentation strategies, which do not generate any extra data, include: (i) Mixup \cite{mixup}, which randomly mixes data batches with corresponding labels; (ii) Random cropping \cite{dcase-resnet}, which randomly crops input features into ones with a smaller length along the time axis; (iii) SpecAugment \cite{spec-aug}, which randomly masks input features by zero along both time and frequency axes. SpecAugment is performed on batch level, and the parameter of mask is set to 10\% of the time and frequency dimensions, respectively.

Strategies that generate extra data include: (i) Spectrum correction \cite{spec-corr}, which aims at transforming a given input spectrum to that of a reference, possibly ideal, device. Different from the original idea, we newly employ spectrum correction as a data augmentation technique, where we create a reference device spectrum, by averaging the spectrum from all training devices except that from device A, and then correct the spectrum of each training waveform collected with device A to obtain extra data. (ii) Reverberation with Dynamic Range Compression (DRC), where we simulate more audio clips from `new devices' by introducing reverberation and applying DRC - this data-agumentation solution is also new for ASC. Room impulse responses \cite{reverb-challenge} are used to create reverberation into the device A data, and DRC is then applied to dynamically compress the amplitude range. Room impulse responses and DRC settings are randomly chosen to generate new training waveforms. (iii) Pitch shift, where we randomly shift the pitch of each audio clip based on the uniform distribution. (iv) Speed change, where we randomly change the audio speed based on the uniform distribution. If output waveform is longer than the original one, extra samples are dropped from the end, otherwise, padding is applied till attaining the same input length. (v) Random noise, we randomly add Gaussian noise to each training waveform. (vi) Mix audios, where we randomly mix two audios from the same acoustic scene class. It's device-independent so it may help simulate a `new device'. Although the solutions from (iii) to (vi) are not new strictly speaking, their application for ASC device robustness has never been evaluated before.

\vspace{-0.2cm}
\section{Experiments \& Analysis}
\label{sec:exp}
\vspace{-0.2cm}

\subsection{Experimental Setup}
\vspace{-0.1cm}
All ASC systems are evaluated on the DCASE 2020 task1a development data set \cite{dcase2020}, which consists of $\sim$14K 10-second single-channel train audio clips and $\sim$3K test audio clips recorded by 9 different devices, including real devices A, B, C, and simulated device s1-s6. Only device A, B, C, s1-s3 are in the training set; whereas, devices s4-s6 are not available in the training phase. The greatest amount of training audio clips are recorded with device A, namely  over 10K audio clips. In the test set, the number of audio clips from each device is the same. For each input audio clip, a short-time Fourier transform (STFT) with 2048 FFT points is applied, using a window size of 2048 samples and a hop length of 1024 samples. The Librosa \cite{librosa} library is employed to extract log-mel filter bank (LMFB) features. Log-mel deltas and delta-deltas without padding are also computed, which finally generates an input feature tensor with size of $423 \times 128 \times 3$. Before feeding the input tensors into the CNN classifier, we perform utterance-level scaling operation to scale LMFB coefficients into [0,1]. All deep models in this work are built with Keras \cite{keras}. Stochastic gradient descent (SGD) with a cosine-decay-restart learning rate scheduler is used to train all deep models. Maximum and minimum learning rates are 0.1, and 1e-5, respectively.\footnote{\scriptsize Code available: \url{https://github.com/MihawkHu/DCASE2020_task1}}

\vspace{-0.1cm}
\subsection{Experimental Results}
\vspace{-0.1cm}
 Table~\ref{tab:3class} shows the  accuracy for 3-class  classifiers,  namely Resnet, FCNN and Ensemble of these two models. The 3-class classification task is relatively easy since an accuracy above 90\%  is attained by each of the three models. FCNN shows a better accuracy than Resent, and the ensemble of these two models allows a further boost of the classification result. In the following experiments with the two-stage system, we use Ensemble system as the 3-class classifier.

\begin{table}[t]
\centering
\caption{Classification accuracy of different 3-class models.}
\label{tab:3class}
\vspace{0.2cm}
\begin{tabular}{l|ccc}
\hline
\hline
3-class Model & Resnet & FCNN & Ensemble \\
\hline
\hline
Acc. \% & 91.4 & 92.9 & 93.2 \\
\hline
\hline
\end{tabular}
\vspace{-0.4cm}
\end{table}

\begin{table}[t]
\centering
\caption{Overall results on DCASE 2020 task1a development set. All proposed data augmentation strategies are used in all our models.}
\label{tab:all}
\vspace{0.2cm}
\begin{tabular}{l|cccc|c}
\hline
\hline
System     & \begin{tabular}[c]{@{}c@{}}A\\ \%\end{tabular} & \begin{tabular}[c]{@{}c@{}}B\&C\\ \%\end{tabular} & \begin{tabular}[c]{@{}c@{}}s1-s3\\ \%\end{tabular} & \begin{tabular}[c]{@{}c@{}}s4-s6\\ \%\end{tabular} & \begin{tabular}[c]{@{}c@{}}Avg.\\ \%\end{tabular} \\
\hline
\hline
Baseline~\cite{dcase2020}  & 70.6                                           & 61.6                                              & 53.3                                               & 44.3                                               & 54.1                                             \\
\hline
Suh et al. \cite{dcase2020-rank1}  &  -                                          & -                                              & -                                               & -                                              & 74.4                                            \\
Gao et al. \cite{dcase2020-rank3}  &  -                                          & -                                              & -                                               & -                                              & 72.5                                           \\
Liu et al. \cite{dcase2020-rank4}  &  -                                          & -                                              & -                                               & -                                              & 72.1                                           \\
Koutini et al. \cite{dcase2020-rank5}  &  -                                          & -                                              & -                                               & -                                              & 73.3                                           \\
\hline
Resnet    &  83.0                                          & 76.1                                              & 73.6                                               & 71.0                                               & 74.6                                             \\
FCNN      & 87.3                                           & 79.5                                              & 75.7                                              & 73.0                                               & 76.9                                             \\
fsFCNN    & 83.9                                           & 78.6                                              & 75.4                                               & 72.8                                               & 76.2                                             \\
Ensemble  & 87.0                                           & 81.5                                              & 78.0                                               & 76.9                                               & 79.4                                             \\
\hline
2-stage Resnet    & 84.5                                           & 78.6                                              & 76.2                                               & 76.4                                               & 77.7                                             \\
2-stage FCNN      & 89.1                                           & 82.9                                              & 78.5                                               & 76.9                                               & 80.1                                             \\
2-stage fsFCNN    & 83.9                                           & 81.2                                              & 78.6                                               & 76.4                                               & 79.0                                             \\
2-stage Ensemble    & \textbf{87.9}                                           & \textbf{84.1}                                              & \textbf{80.4}                                              & \textbf{79.9}                                               & \textbf{81.9}                                             \\
\hline
\hline
\end{tabular}
\vspace{-0.5cm}
\end{table}

All main experimental results concerning the 10-class ASC task are shown in Table~\ref{tab:all}. In the training set, device A data accounts for around 75\%, and device B, C, s1-s3 accounts for around 5\%, respectively. Thus, we can think of device A as a source device, and the remaining ones as target devices. Based on the device information of data, we divide the test set into four different subsets, which represent real source data (device A), real target data (device B \& C),  target seen-simulated data (device s1-s3), and target unseen-simulated data (device s4-s6). The first row in Table~\ref{tab:all} gives ASC accuracy of the official baseline system \cite{dcase2020}, which uses a two-fully connected layer neural classifier, and OpenL3 \cite{openl3} to extract input audio embedding. From baseline results, we can argue that good ASC accuracy can be attained on real source data (device A), but a severe performance drop is observed on other devices. Specifically, a severe ASC accuracy drop is reported on unseen devices (s4-s6), where the ASC accuracy is as low as 44.3\%. Rows from 6th through 9th in Table~\ref{tab:all} display ASC results attained with a 10-class classifier. Resnet attains a significant improvement on all test devices over the baseline system.  FCNN and fsFCNN  outperform  Resnet on all testing scenarios, and FCNN achieves slightly better results than fsFCNN. Finally, the ensemble of these three models attains a 79.4\% overall ASC accuracy, which is a meaningful improvement compared to any of the single-stage models reported in Table~\ref{tab:all} .

The experimental results of two-stage systems are shown in the last four rows of Table~\ref{tab:all}. When comparing single-stage 10-class classifier with the two-stage system, a significant improvement can be attained. For example, comparing `Resnet' with `2-stage Resnet`, 3.1\% absolute improvement on overall accuracy is observed (74.6\% vs. 77.7\%). Accuracies on all the test sets have indeed improved, especially on the unseen-simulated test set (device s4-s6), where a 5.4\% absolute improvement is obtains. The overall absolute accuracy gains with the proposed two-stage technique are 3.2\% for FCNN, 2.8\% for fsFCNN, and 2.5\% for the ensemble. The best performance is attained by the two-stage of three models' ensemble, which achieves 81.9\% overall ASC accuracy. Although the top performance is still delivered on  the source device A (87.9\%), the average ASC accuracy on the remaining test devices is above 79\%, which compares favourably against the baseline system. When comparing ASC accuracy between seen-simulated test set (device s1-s3) and unseen-simulated test set (device s4-s6), we can argue that the gap in performance is mild (80.4\% vs. 79.9\%), and they are close to the overall average accuracy (80.4\% vs. 79.9\% vs. 81.9\%). Those results confirm the effectiveness of the proposed two-stage system with respect to improving device robustness.

To better compare our results, the results of top-5 ranking systems (excluding ours) \cite{dcase2020-rank1, dcase2020-rank3, dcase2020-rank4, dcase2020-rank5} are also listed from 2st to 5th rows in Table~\ref{tab:all}. The experimental results clearly demonstrated that the proposed two-stage system achieves a competitive performance with all benchmark models.

\begin{table}[t]
\centering
\caption{Accuracy comparison of different data augmentation strategies on Resnet. Mixup and random cropping are always employed. `sa' indicates specAugment. `sc' indicates spectrum correction. `r' indicates reverberation with DRC. `aug' indicates another four augmentation methods, including pitch shift, speed change, random noise and mix audios.}
\label{tab:resnet}
\vspace{0.2cm}
\begin{tabular}{l|cccc|c}
\hline
\hline
System      & \begin{tabular}[c]{@{}c@{}}A\\ \%\end{tabular} & \begin{tabular}[c]{@{}c@{}}B\&C\\ \%\end{tabular} & \begin{tabular}[c]{@{}c@{}}s1-s3\\ \%\end{tabular} & \begin{tabular}[c]{@{}c@{}}s4-s6\\ \%\end{tabular} & \begin{tabular}[c]{@{}c@{}}Avg.\\ \%\end{tabular} \\
\hline
\hline
Resnet         & 78.8                                           & 72.1                                              & 69.3                                               & 69.5                                               & 71.0                                             \\
\ \ +sa      & 80.3                                           & 73.5                                              & 71.4                                               & 67.7                                               & 71.6                                             \\
\ \ +sa+sc   & 79.1                                           & 75.0                                              & 70.7                                               & 68.9                                               & 72.0                                             \\
\ \ +sa+sc+r & 80.3                                           & 74.7                                              & 71.4                                               & 70.3                                               & 72.8                                            \\
\ \ +sa+sc+r+aug$^*$ & 83.0                                          & 76.1                                              & 73.6                                               & 71.0                                             & 74.6                                       \\
\hline
\hline
\end{tabular}
\vspace{-0.5cm}
\end{table}

\vspace{-0.1cm}
\subsection{Evaluation of Data Augmentation Strategies}
\vspace{-0.1cm}
We now investigate the effectiveness of different data augmentation strategies in combination. We test only the Resnet-based ASC system to keep the experimental setup consistent. Table~\ref{tab:resnet} shows the classification accuracy attained with different data augmentation schemes. It should be noted that mixup and random cropping are always employed in the training phase.

When comparing the first two rows, we can see that specAugment can boost the overall ASC accuracy from 71.0\% to 71.6\%, but  on unseen data the accuracy becomes worse (from 69.5\% to 67.7\%). Next, we add spectrum correction, which results in  an increase in the amount of training data. As described before, device A data is used to generate data from other devices, in which we get the same size extra data with device A data. From the results in Table~\ref{tab:resnet}, we can observe that spectrum correction has a beneficial effect on overall classification results, but the accuracy on device A decreases. That outcome is not unexpected  since there is now much more training data from other devices than from device A. New extra data is generated by reverberation with DRC, which aims to simulate data from more different devices. From the results we can see when extra data with reverberation and DRC is added in the training set, the overall ASC accuracy is further boosted (72.8\%), where most of the gain is imputable to an increse of the accuracy on the unseen test data. Finally, we employ four more data augmentation approaches, including pitch shift, speech change, random noise and mix audios. As shown in the last row of Table~\ref{tab:resnet}, performance can be further boosted. Using all strategies, a 74.6\% accuracy on Resnet is attained\footnote{We also increased amount of Resnet parameter, i.e, the channel number, when all data-augmentation schemes are used.}.

\begin{figure}[t]
  \centering
  \includegraphics[width=.9\linewidth]{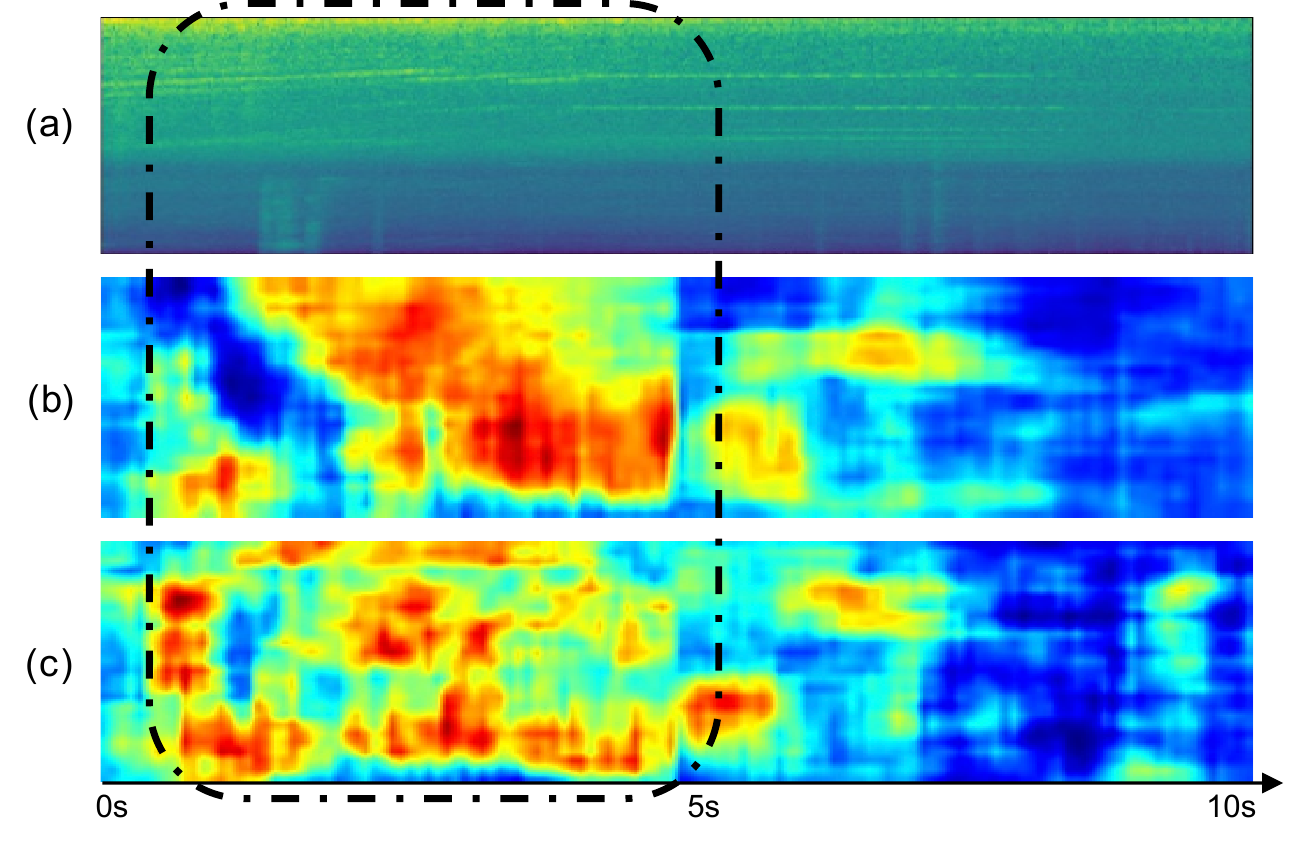}
  \vspace{-0.4cm}
  \caption{Nerual saliency analysis via CAM of an audio clip from a metro station ($metro\_station\_vienna\_87\_2389\_a.wav$). The three plots are: (a) spectrogram, (b) CAM of 3-class classifier, (c) CAM of 10-class classifier. In this 10-second audio, brake and horn sound starts from 0s to around 8s and only reverberation remains after 5s.
}
  \label{fig:cam1}
     \vspace{-0.4cm}
\end{figure}

\vspace{-0.2cm}
\subsection{Neural Saliency Analysis via Class Activation Mapping}
\vspace{-0.1cm}
Neural saliency methods~\cite{cam, owens2018audio} aim to provide interpretable analyses on the weight distribution over hidden neurons of a well-trained artificial neural network model. Benchmark saliency methods, such as class activation mapping (CAM)~\cite{cam}, have shown a correlation between model performance and ability in spotting physical pattern in DNN-based  acoustic  models~\cite{owens2018audio}. 

\begin{figure}[t]
  \centering
  \includegraphics[width=.9\linewidth]{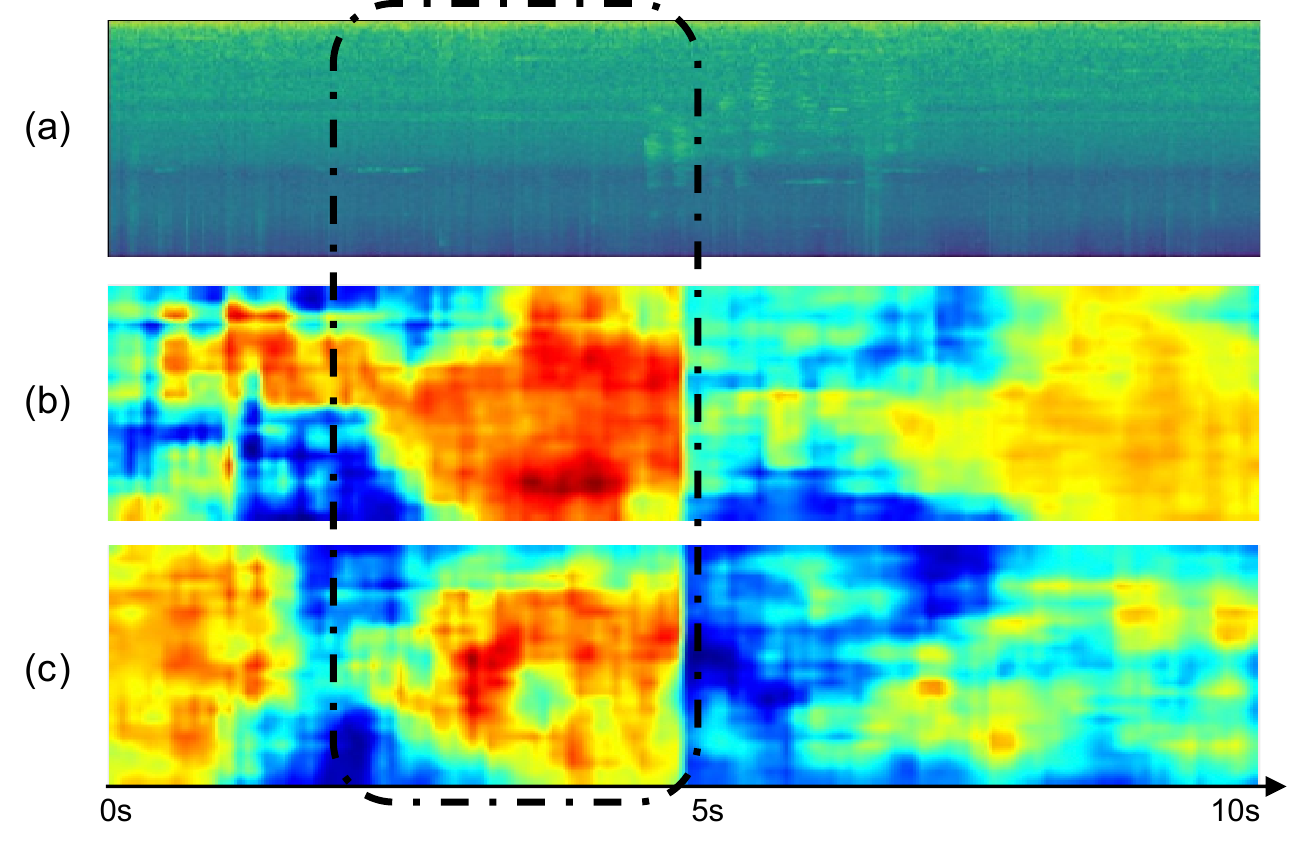}
  \vspace{-0.4cm}
  \caption{Nerual saliency analysis via CAM of an audio clip from a bus ($bus\_prague\_1102\_42431\_a.wav$). The three plots are (a) spectrogram, (b) CAM of 3-class classifier, (c) CAM of 10-class classifier, respectively. In this 10-second audio, brake sound starts from around 2s to around 5s, human talk starts from around 5s to 7s.}
  \label{fig:cam2}
   \vspace{-0.4cm}
\end{figure}

In this work,  we use CAM \cite{cam} to gain a better understanding about what  sound patterns are found by our CNNs to accomplish the ASC task. Indeed, CAM can highlight the class-specific discriminative regions in the input feature map. In other words, CAM helps generating a two-dimension activation map that can be used to better interpret the prediction decision made by deep architectures. In our case, where an audio clip is transformed into a time-frequency representation, CAM reveals whether the ASC decision is triggered by CNN's attention posed on time and frequency regions of input audio clip having a meaningful semantic content with respect to the target acoustic scene. Examples of some analysis results obtained with CAM are shown in Figures~\ref{fig:cam1} and \ref{fig:cam2}, where the spectrogram of the input audio clip (top panel), and the CAM for the 3-class classifier (center panel), and the 10-class classifier (bottom panel) are shown. Regions in CAM with deep red color imply that the CNN pays more attention to them, and those regions have a higher class-specific discriminative power. As the size of FCNN and fsFCNN's output feature map is too small to map back to the input, only Resnet model is used for neural saliency analysis.

In Figure~\ref{fig:cam1}, a 10-second audio clip referring to a metro station scene is visualized; the audio clip contains brake and horn sounds from 0s to around 5s. After 5s, only reverberation remains in the audio. The CAM result indeed reveals that the CNN pays more attention to the segment between 0s and 5s.  Moreover, 3-class classifier and 10-class classifier have similar activation regions, and the 3-class classifier seems to give more weight to the information in that region. Indeed, if we check the confidence in making the prediction, i.e., scaled class posterior probability from the CNN,  the proposed two-stage fusion can increase it from 70.7\% to 85.0\%.  The same pattern can be observed in the results of Figure~\ref{fig:cam2}, in which it contains a 10-second audio from a bus station. This audio has brake sound between around 2s and 5s, and human speech starts around 5s till 7s. From the CAM results, we can see that CNN pays more attention to the segment from 2s to 5s, which contains a brake sound. However, the CNN is not interested in the human speech although it is clearly audible and related to the bus arriving announcement. This result is indeed interesting: Human beings would use the bus-station announcement to make ASC decision; however, the CNN cannot perform speech recognition, so it may simply classify that region as belonging to an ideal human speech class, which however appears in many scenes and has therefore a low discriminative power. 

We can thus argue that CNNs use acoustic events to classify the input scene audios. Specifically, CNNs pay more attention to audio segments containing acoustic events, such as birds sound from park, or car horn sound from street traffic. It should be noted that our conclusion with CAM is different from that in \cite{asc-cam}, where the authors claim that CNNs use background sound rather than acoustic events. By comparing the experimental setups, the main difference is that each audio clip is chunked into 10 1-second segments in their setup; in contrast, we use the whole 10-second audio clip as the input. Since acoustic events usually last for several seconds, it's difficult for CNNs to capture them when inputs are very short.

\vspace{-0.1cm}
\section{Conclusion}
\label{sec:con}
\vspace{-0.2cm}
This work focuses on device robustness in acoustic scene classification. We propose a novel two-stage ASC framework based on CNNs. Three different fully CNN models and several novel data augmentation strategies are investigated to improve device robustness. A general 3-class classifier and a specific 10-class classifier are combined through score fusion. Experiments on the DCASE 2020 task1a development set show the effectiveness of our solution. Specifically, our best system, a two-stage fusion of a CNN-based ensemble, obtains a state-of-the-art 81.9\% average ASC accuracy. Moreover, our CAM-based neural saliency analysis demonstrates that CNNs pay particular attention to audio segments strictly related acoustic events rather than simply fetching information from the background environmental sound.

\clearpage

\bibliographystyle{IEEEbib}
\bibliography{refs}

\end{document}